\documentclass[conference]{IEEEtran}
\usepackage{graphicx}
\usepackage[utf8]{inputenc}
\ifCLASSINFOpdf
\else
\fi

\usepackage{hyperref} 
\hypersetup{
     hidelinks
}

\begin{document}
\title{Hardened Paxos Through Consistency Validation}

\author{\IEEEauthorblockN{Rodrigo R. Barbieri}
\IEEEauthorblockA{Flextronics Instituto de Tecnologia\\
Sorocaba, São Paulo, Brazil\\
Email: rodrigo.barbieri2010@gmail.com}
\and
\IEEEauthorblockN{Gustavo M. D. Vieira}
\IEEEauthorblockA{Department of Computing at Sorocaba (DComp)\\
Federal University of São Carlos\\
Sorocaba, São Paulo, Brazil\\
Email: gdvieira@ufscar.br}
}


%Telephone: (800) 555--1212, Fax: (888) 555--1212}


\maketitle

\begin{abstract}
%\boldmath
    Due to the emergent adoption of distributed systems when building applications, demand for reliability and availability has increased. These properties can be achieved through replication techniques using middleware algorithms that must be capable of tolerating faults. Certain faults such as arbitrary faults, however, may be more difficult to tolerate, resulting in more complex and resource intensive algorithms that end up being not so practical to use. We propose and experiment with the use of consistency validation techniques to harden a benign fault-tolerant Paxos, thus being able to detect and tolerate non-malicious arbitrary faults.

\end{abstract}


% \fi
\IEEEpeerreviewmaketitle

\section{Introduction}

Distributed systems have often been used as a basis for a wide range of services and applications. The adoption of this model is motivated by the need for satisfying requirements that become indispensable as we become dependent on automated systems, requirements such as reliability and availability. It is quite difficult to guarantee these two properties in distributed systems due to the possibility of partial failure of the involved components. One of the most popular approaches to improve a distributed system's availability is replicating the application~\cite{faultTolerance,bookIntroduction}, so failures do not compromise availability because there are several replicas providing the same application. Reliability is more difficult to guarantee, because the replicated application must not have its state or part of it corrupted when facing a failure.

There are several factors that make it more difficult to synchronize the application state between replicas. The easiest to diagnose causes usually are communication problems, crashes or power failures, that prevent a replica from being updated. However, there are other causes that may be much harder to diagnose, such as data corruption, that may cause the replica to display erroneous behavior and incorrect results. It is useful to classify those faults into two main classes: benign faults and arbitrary faults~\cite{faultTolerance,bookIntroduction}. The first class represents faults that are related to the use of software and hardware components that may stop working at any given time, but they do not deviate from expected behavior. As for the second class, it represents any fault where the components may display any type of behavior, including behavior caused by external malicious attackers.

One of the most well known techniques to implement replication in a consistent way is through active replication. In this model the application can report to the client application that the operation has succeeded only when a minimum number of replicas has committed the change~\cite{schneider,faultTolerance,bookIntroduction}. The minimum number of replicas varies according to the algorithm used and the fault model tolerated. Active replication has been the focus of research when developing middleware for distributed applications. Several solutions such as Paxos~\cite{lamport}, leases~\cite{chubby} and viewstamp~\cite{viewstamp} are implemented in active replication middleware with each its particular approach~\cite{viveladifference}.

Paxos~\cite{lamport} supports benign faults in the crash-recovery model~\cite{bookIntroduction} , where a replica stops working for a certain period of time and later is able to recover itself. In order to accomplish this, Paxos uses persistent memory to save its state and recover in the event of a crash. Paxos guarantees that subsequent updates will force the replica to update its state. Even though Paxos is perfectly tolerant to benign faults, some of the faults mentioned previously are able to compromise its reliability and availability properties. One way of dealing with these faults is hardening the fault model, changing the algorithm appropriately. 

A Paxos algorithm adapted for the arbitrary faults class~\cite{refinement,castroPracticalByzantine} is much more costly than its counterpart for the benign faults class, for the following reasons:

\begin{enumerate}
\item It performs more message exchanges and disk operations;
\item It requires that no more than a third of replicas fail;
\item Its usage of encryption increases system’s overhead~\cite{datacenterNonMalicious};
\item Its implementation is complex~\cite{datacenterNonMalicious,nonMalicious};
\item It requires deployment of different platform and applications versions (heterogeneity)~\cite{hardening,datacenterNonMalicious};
\item The increase in number of replicas reduces its performance and may cause an increase in error occurrence~\cite{hardening,nonMalicious}.
\end{enumerate}

Among the faults tolerated by the algorithm in the arbitrary faults class, there are several types of faults that are not malicious and are relatively common to distributed applications~\cite{paxosMadeLive,hardening}, such as:

\begin{enumerate}
\item Network failures: data corruption during transmission;
\item Hardware failures: corrupt read/write operations on main memory or secondary storage;
\item Programmer failure: error in algorithm implementation;
\item Operator failure: erroneous behavior due to incorrect configuration.
\end{enumerate}

It is possible to improve a benign algorithm to tolerate the previously mentioned faults, and by choosing to not tolerate malicious faults it is possible to achieve a non-malicious arbitrary fault tolerant algorithm that may be less costly than the arbitrary one, as presented in \cite{nonMalicious,hardening}. In this paper we show how to harden the benign crash-recovery Paxos through consistency validation techniques used to detect non-malicious arbitrary faults, like redundancy~\cite{nonMalicious}, integrity~\cite{hardening} and semantic~\cite{datacenterNonMalicious} validations. Our main contribution is that we apply these strategies at the middleware level, allowing any application built on top of the middleware to be automatically hardened.

In order to implement, test and validate this work, we use a Paxos-based Java library known as Treplica~\cite{vieira08a,vieira-tr10b}. We used fault injection techniques to test and validate the implementation. We were able to harden the originally implemented benign crash-recovery fault model to a crash-stop non-malicious arbitrary fault model by being able to detect the previously mentioned arbitrary faults. The end result is that any application built on top of the hardened Treplica is crash-stop non-malicious arbitrary fault tolerant. 

The remainder of the paper is organized as follows. In Section II we describe fault models and Paxos under benign fault class. In Section III we describe some arbitrary fault class consensus algorithms and their limitations. In Section IV we describe our approach and in Section V we present the results. We talk about related work in Section VI and finally conclude in Section VII.

\section{Replication under benign faults}

In distributed systems we use the concept of fault models to abstract the properties a system must satisfy and which faults a distributed algorithm for this system must tolerate. For the benign faults class, there are two classic fault models~\cite{bookIntroduction,schneider} that draws our attention: 

\begin{enumerate}
\item crash-stop: replicas that fail are removed from the algorithm permanently;
\item crash-recovery: replicas that fail can recover and continue to participate in the algorithm;
\end{enumerate}

The fault models above range from weaker (more strict) to stronger (more general). The stronger the model, the more complex and difficult it is to implement an algorithm. When building a practical distributed system, it is desirable to adopt a fault model that better fits the system and satisfies its requirements for performance and types of faults it must tolerate, but this is not always the case, since any distributed system that relies on actual computers is prone to arbitrary faults.

\subsection{Active Replication}

Distributed applications built on top of active replication middleware are modeled as deterministic finite state machines. Each system operation is modeled as a state transition, where a state consists in a set of information that includes the previous state and a transition to the current state. Each replica is a state machine on its own and the algorithm makes use of total order broadcast to propagate the transitions in a consistent way. Ultimately, all replicas are kept synchronized in the same state because their transition messages are processed in the same order~\cite{bookIntroduction,schneider}, using an atomic broadcast or consensus algorithm.

\subsection{Paxos}

Paxos is a consensus-based active replication algorithm proposed for asynchronous systems augmented with failure detectors~\cite{bookIntroduction}. Paxos also assumes a crash-recovery fault model that tolerates benign faults~\cite{lamport}. Replicas agree on a certain value or operation through voting, the decision determines what operation is executed on all replicas. A voting round consensus is reached when the coordinator receives successful votes from the majority of replicas, then the decision is broadcast.

In order to satisfy the reliability and availability properties, Paxos must recover its state when a replica fails. Its approach to accomplish this is to save a log of all its operations in persistent memory, including proposals, votes and decisions, so when the application is restarted, it is able to replay the log and get back to the state prior to the crash.

\section{Replication under arbitrary faults}

According to literature~\cite{bookIntroduction}, the fault models for arbitrary faults class include:

\begin{enumerate}
\setcounter{enumi}{2}
\item fail-arbitrary: similar to crash-recovery, but replicas must tolerate any failure (benign and arbitrary);
\end{enumerate}

Many algorithms solve consensus (and active replication) in the fail-arbitrary fault model. One of the first was described as “The Byzantine Generals Problem”~\cite{generals}. This paper studies the problem where war generals try to reach an agreement while one of them is a traitor. Due to this work, the fault model is often referred to as Byzantine faults.

In the proposed algorithm, several rounds of voting through encrypted message exchanges are required to detect malicious replicas, where the quorum size increased from the single majority required from the crash-recovery model to more than two thirds of replicas to ensure consensus. The first practical Byzantine algorithm used Paxos in the crash-stop model~\cite{schneider}, and later a more robust solution was published~\cite{castroPracticalByzantine}, where the crash-recovery Paxos is able to tolerate arbitrary faults. This more robust solution made use of Message Authentication Codes (MACs), real-time assumptions, communication protocol through the file-system, and optimizations such as reconfiguration, garbage collection and state transfer.

Lamport analyzed the algorithm in \cite{castroPracticalByzantine} and proposed in \cite{refinement} a more general version of the algorithm, derived directly from Paxos. The new more general algorithm makes use of digital signatures and is changed so that all replicas exchange proposal, vote and decision messages. Although it solves the “Byzantine Generals Problem”, its increased performance impact and the fact that malicious faults are being tolerated through solutions orthogonal to distributed systems middlewares~\cite{datacenterNonMalicious,hardening}, contributed to its low-adoption in practical distributed systems. Byzantine Paxos has also been criticized for the following problems:
\begin{enumerate}
\item Although faults are tolerated, it may not be possible to find out the cause~\cite{datacenterNonMalicious,hardening};
\item If there is no heterogeneity in the system, operator and programmer faults cannot be detected~\cite{hardening,datacenterNonMalicious};
\item Increasing the number of replicas reduces performance and possibly increases fault incidence~\cite{hardening,nonMalicious};
\item If more than a third of replicas fail, it is not possible to detect that the system has been compromised~\cite{schneider}.
\end{enumerate}

\section{Non-malicious arbitrary Paxos}

\begin{figure}[htbp]
\centering
\includegraphics[width=0.48\textwidth]{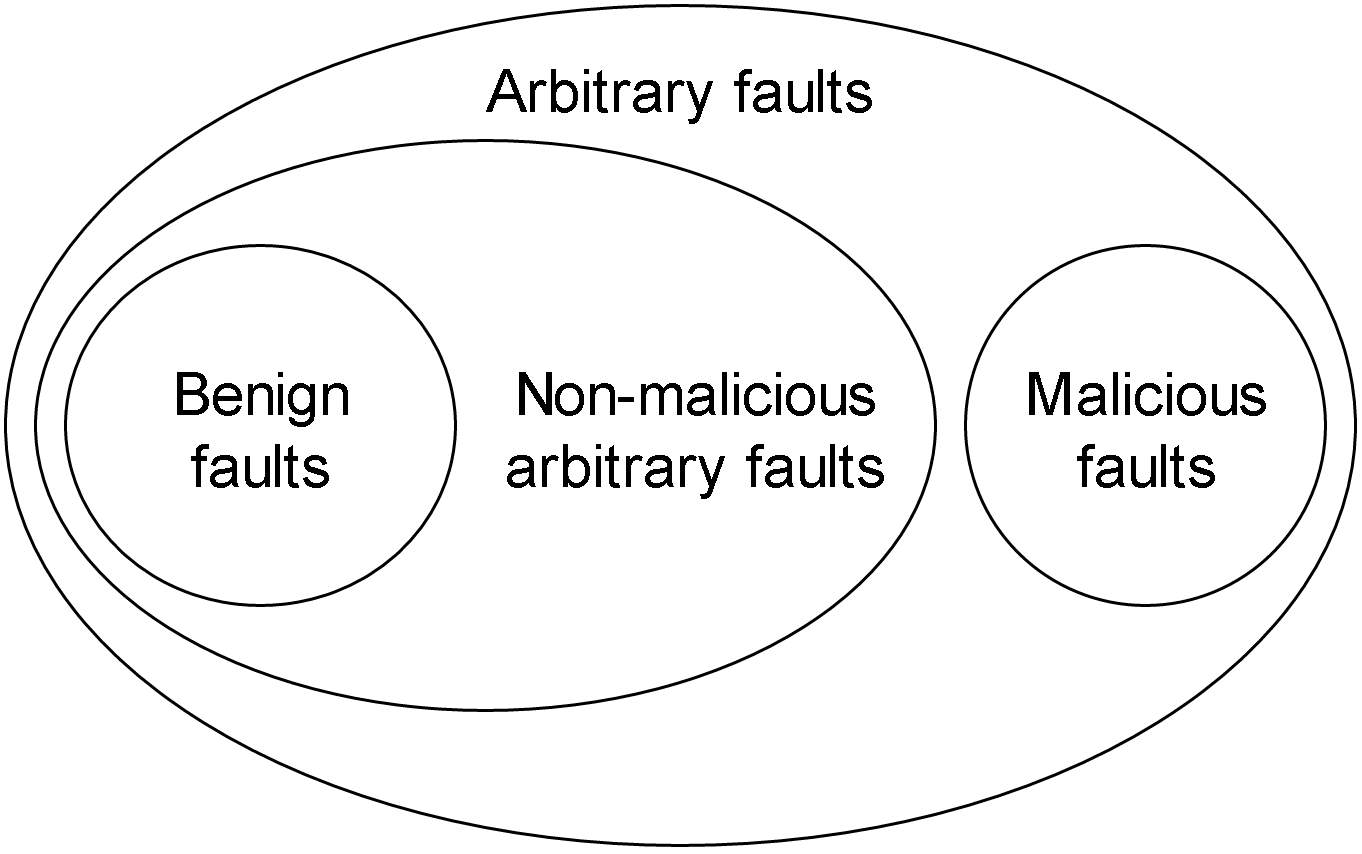}
\caption{Representation of non-malicious arbitrary faults class amidst other classes.}
\label{fig:models}
\end{figure}

Many practical distributed system implementations desire to tolerate arbitrary faults, but would prefer a less performance intensive algorithm than a byzantine one~\cite{datacenterNonMalicious,hardening,nonMalicious}. While malicious faults are being tolerated using different techniques~\cite{datacenterNonMalicious,hardening}, and based on the premise that any fault model can be hardened to tolerate some arbitrary faults, it is possible to harden the crash-recovery benign model to tolerate non-malicious arbitrary faults, thus achieving the following fault model, as shown in Figure~\ref{fig:models}:

\begin{enumerate}
\setcounter{enumi}{3}
\item fail-arbitrary non-malicious: similar to fail-arbitrary, but malicious faults are not tolerated by the algorithm.
\end{enumerate}

An algorithm for the above fault model is considered to be less complex than a fail-arbitrary one for not tolerating malicious faults in its implementation. However, the implementation required to tolerate all non-malicious arbitrary faults adds its own complexity to the algorithm.

\subsection{Consistency Validation}

Non-malicious arbitrary fault types are present not only in any practical distributed system, but in any system that relies on computer components. These faults are tolerated through various approaches, such as:

\begin{enumerate}
\item Data and state redundancy~\cite{datacenterNonMalicious,hardening,scrubbing,memoryIntegrity}: each process variable or stored data has a duplicate which can be used for validation and backup;
\item Checksums and hashes~\cite{datacenterNonMalicious,hardening,memoryIntegrity,scrubbing}: generating a checksum or hash of data and attaching it to the protocol messages prior to transmitting across the network or saving them in storage allows for future detection of undesired corruption;
\item Integrity checks~\cite{datacenterNonMalicious,hardening,memoryIntegrity,scrubbing}: validate data integrity on each read and write operation, usually making use of redundant data;
\item Semantic checks~\cite{datacenterNonMalicious}: validate that after an operation has been applied on data, the newly obtained state is semantically correct according to the applied operation. For instance: after adding an element to a list, check if the element is in the list;
\item Arithmetic codes~\cite{nonMalicious}: numerical properties of data are used to detect undesired corruption. For instance: If numerical variables are multiplied by a prime number upon writing and divided by the same number when they are read back, the remainder should always be zero. 
\end{enumerate}

Each approach mentioned above has its overhead cost associated, for either performing repeated checks, encrypting, or doubling memory requirements due to redundancy. We decided to create a unique set of validation techniques to harden the Paxos algorithm, looking for the ones that best match the software architecture of the middleware used.

\subsection{Treplica}

Treplica~\cite{vieira08a,vieira-tr10b} is a library coded in Java that allows distributed applications to use Paxos as middleware to manage state replication through its state machine. Its implementation is very close to a traditional Paxos~\cite{lamport} implementation. In Treplica, replicas can assume any Paxos role, such as Coordinator, Proposer, Learner and Acceptor. Applications designed according to the Model-View-Controller standard can easily be modeled to use Treplica. 
We chose Treplica because its modular architecture allows for improvements to be easily coded and tested. Since it is designed to tolerate benign faults, upon analysis we validated that it is prone to non-malicious arbitrary faults we are interested in, due to:
\begin{enumerate}
\item Reading and Writing serialized binary files to the storage, which can become corrupt on storage failure;
\item Usage of UDP protocol for message exchanges, which can be corrupted on noisy network channels; 
\item Usage of Java virtual machine, which can have its process memory space corrupted at runtime.
\end{enumerate}

Additionally, Treplica is object oriented and makes use of immutable objects design, where an object is never changed after being instantiated. This allows for more efficient use of checksums. State transition semantic checks can also be easily coded by the application due to its integration with the state machine modelling.

\subsection{Hardening the benign crash-recovery fault model}

Our main approach to harden our existing benign crash-recovery fault model towards the fail-arbitrary non-malicious one is to employ consistency validation checks to detect arbitrary faults, while initially not worrying about how to recover from them. From the point of view of a benign fault model distributed system, most arbitrary faults behave as silent faults because they cannot be detected. For instance, if a user clicks a button to buy one book, but a replica processes that two books have been bought because bits got flipped along the way, then this is not a fault from Paxos point of view, because the message was delivered consistently across all replicas. In order to effectively detect such silent faults, we employed the following techniques:

\subsubsection{Data corruption detection}
We often check for data corruption as soon as it can be detected. Whenever an immutable Paxos message object is instantiated, either to be written to persistent storage or propagated to the network, we calculate a checksum of its contents and append to it. When the message is received or recovered from storage, the checksum is recalculated and validated by comparing it to the one attached. We acknowledge that recalculating a hash every time some data is read adds overhead, but Treplica’s modular architecture allowed us to identify key locations in the source code to minimize overhead. Through this implementation we were able to detect any corruption that affects messages, such as network messages, persistent storage and main memory corruption.

\subsubsection{State integrity checks}
We also attempt to detect application state corruption by having a duplicate state and by maintaining a current state checksum. Every time a state transition takes place, we apply the transition operation to both states, then we call a method implemented by the application that should return an object that best describes its current state. This allows for any transition operation that silently fails and causes the states to diverge to be detected before any further harm is done. Both states are also validated every time the application state object is requested, since they can become corrupt any time due to memory corruption. The current state checksum is recalculated after each successful state transition based on the object's data and the previous checksum. This checksum allows for state integrity validation between replicas, either by including it in the protocol or opportunistic messages.

\subsubsection{Semantic checks}
We introduced semantic validation that helps detect main memory corruption and programmer errors. For each state transition operation implemented by the application, it is enforced to implement a semantic validation method that verifies if the transition has been correctly applied to the state. This semantic validation method is run as soon as the state transition is applied, thus if the validation fails, all further operations on the given replica are halted.

\section{Experimental Validation}

The main purpose of testing our implementation is to validate whether the hardened Treplica is more fault tolerant than the unmodified Treplica. By injecting faults in the unmodified Treplica, we could observe the behavior caused by them, such as crashes, lock-ups and erroneous behavior, which properties they compromise and which faults the hardened Treplica should be tolerant to. 

\subsection{Fault Injection}

In order to test our implementation through fault injection, we used an aspect-oriented library, known as AspectJ. It allows us to change the behavior of any Java program without changing its main code. Our technique was to generate corruption faults on the message protocols and state transitions through byte flips and value changes that would attempt to pass undetected through our validation. The same fault injection code is also compatible with the unmodified Treplica, so we could easily compare it to our hardened version. We coded all fault injections separately and added probability settings, which allows us to control the rate at which each fault is injected. 

\begin{figure}[htbp]
\centering
\includegraphics[width=0.48\textwidth]{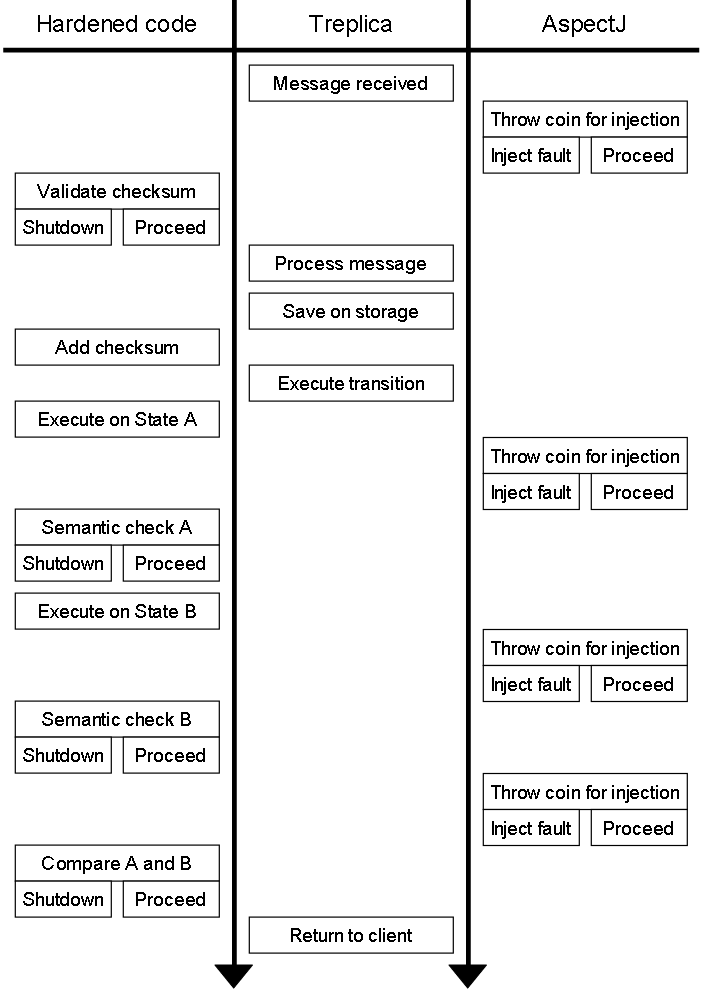}
\caption{Experiment code work flow running with fault injection in Treplica.}
\label{fig:codeflow}
\end{figure}

For protocol messages, we change a random value in the message as soon as the message is received from the network or recovered from storage, while retaining the checksum value on the hardened Treplica. The protocol message fault injections are then detected before processing the message. We created an example application built on top of Treplica, that consists of a list of strings. We injected faults in the state by either removing or adding elements, and by changing the string themselves. The application method that returns the state description was able to capture the differences caused by the fault injections on the following transition, by comparing to the duplicate state. Transition fault injections consisted of not doing the transition operation, which were also captured by the duplicate states and subsequently by the semantic checks. The resulting code work flow for the experiment execution on a replica can be seen in the diagram shown in Figure~\ref{fig:codeflow}.

\subsection{Analysis}

By injecting the above faults into the original Treplica, we were able to confirm that reliability and availability properties were compromised. For protocol message injections, the replicas would not only display benign faults such as lock-ups and crashes, but also perform incorrect application operations, vote in incorrect rounds, and flood the network with invalid round messages. Other faults injected did not affect Paxos, but they significantly affected the application state consistency, noticeable by the client. Our hardened Treplica however, was able to detect all the faults injected. Through fault injection we were also able to detect bugs in our implementation that did not appear during our regular testing.

Diagnosing faults was a challenge on its own. A storage corruption resulted on the same replica failing over and over because it was reading corrupt data every time it restarted. A corrupt network message can be dropped, thus becoming a benign failure. Main memory corruption was found to be the most difficult to cover, test and diagnose.

Our fault detection techniques on messages received and recovered prevent a replica from processing and further spreading a fault that originated in a different replica or in storage. We can list two types of faults that could still be propagated:

\begin{enumerate}
\item The highly unlikely case of data corruption between instantiating the immutable object and generating its checksum (or generating an incorrect checksum), for this scenario we are not taking any action;
\item Main memory corruption in internal Paxos state, which can lead to erroneous behavior, like a replica getting lost between voting rounds, voting incorrectly or the Coordinator starting invalid voting rounds. Although not yet implemented, our analysis indicates that using hashes for this type of validation would degrade system performance greatly, thus we are considering a different approach, in which replicas would analyze each other’s messages received to detect erroneous behavior.
\end{enumerate}

\begin{figure}[htbp]
\centering
\includegraphics[width=0.48\textwidth]{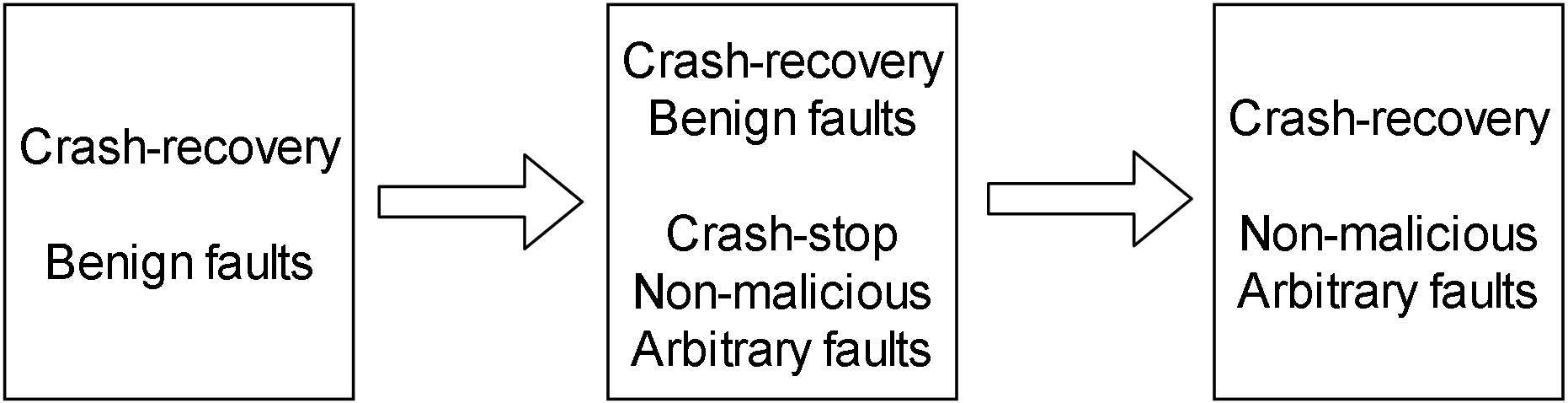}
\caption{Hardening plan starting from crash-recovery towards crash-recovery non-malicious arbitrary fault model.}
\label{fig:hardening}
\end{figure}

According to our analysis, we believe that upon detecting the faults, the most effective approach to recover a replica from most arbitrary faults is to transfer a fault-free state from another replica, resetting the replica to a pristine state in the distributed system’s state machine. However, our solution so far has been to shutdown the replica instance because Treplica currently does not have a state transfer feature implemented. This solution would result in a crash-stop non-malicious arbitrary fault model instead of the one we initially intended to achieve. There is only a limited number of faults we can recover from while we do not have state transfer feature. Figure~\ref{fig:hardening} displays what fault tolerance we achieved in our experiment.

We consider the crash-stop non-malicious arbitrary fault model to be more resilient and more practical than the original crash-recovery implementation. If a benign fault occurs, the system is able to recover itself and continue, but if a non-malicious arbitrary fault is detected and is non-recoverable, we shutdown the replica, preventing any propagation of erroneous behavior.

\section{Related Work}

Other research has been published trying to achieve arbitrary non-malicious fault tolerance through several different approaches. The most common approach observed is to harden a less tolerant fault model towards the most tolerant one by covering each type of fault present in the arbitrary fault class individually. It also seems to be the case where the use of hashes or Message Authentication Codes (MACs) to provide validation checks is one of the most plausible solution to several types of faults. 

In \cite{hardening}, an in-depth non-malicious arbitrary faults study is presented. Many of this paper techniques are inspired by this work. The approach taken was to develop a library that hardens processes built on top of it. All the process' messages, event handlers and variables, if implemented according to the library, are managed by it as part of its state. It intercepts all messages and event handlers to perform integrity checks on them, and aborts whenever a fault is detected. This library is not a middleware, but it can be used to harden existing benign fault tolerant middlewares if implemented on top of the library. Our approach takes an existing middleware and explores the challenges of hardening the middleware itself.

In \cite{datacenterNonMalicious}, although it discusses several concepts on detecting arbitrary faults, it only implements semantic checks. This is similar to part of our approach, with a low coverage because the checks are implemented only at the application layer.

The approach presented in \cite{nonMalicious} involves the use of a low-level encoding compiler so processes read, write and perform all operations with encoded arithmetic values. Whenever a value is changed due to corruption, the arithmetic decoding operation fails and process detects it. Arbitrary faults handling is mapped to benign faults, so processes either crash or have their messages discarded. This approach also sacrifices error coverage for better performance due to the use of arithmetic codes.

\section{Conclusion}

Among the fault models used to build distributed systems, crash-recovery and fail-arbitrary stand out for benign and arbitrary fault classes, respectively. There is a big difference in types of faults tolerated and also in resource requirements for each of those fault models, where fail-arbitrary has not been the preferred model. It is possible to propose a set of consistency validation techniques that allows benign crash-recovery algorithms to be hardened towards the same coverage as fail-arbitrary algorithms while not considering malicious faults. By implementing and experimenting with those techniques on a Paxos-based library, we hardened our fault model, successfully detecting non-malicious arbitrary faults and achieving a crash-stop non-malicious arbitrary fault model by shutting down the replica once a fault is detected. Our work currently does not recover from such faults, but at this point, we consider the crash-stop non-malicious arbitrary fault model to be more resilient and more practical than fail-arbitrary, also requiring less effort on developers to create a fault tolerant application for this fault model using such middleware.

% or

%\appendices
%\blindtext

%\section*{Acknowledgment}


\ifCLASSOPTIONcaptionsoff
  \newpage
\fi



\bibliographystyle{IEEEtran}
\bibliography{IEEEabrv,main}

\begin{thebibliography}{10}
\providecommand{\url}[1]{#1}
\csname url@samestyle\endcsname
\providecommand{\newblock}{\relax}
\providecommand{\bibinfo}[2]{#2}
\providecommand{\BIBentrySTDinterwordspacing}{\spaceskip=0pt\relax}
\providecommand{\BIBentryALTinterwordstretchfactor}{4}
\providecommand{\BIBentryALTinterwordspacing}{\spaceskip=\fontdimen2\font plus
\BIBentryALTinterwordstretchfactor\fontdimen3\font minus
  \fontdimen4\font\relax}
\providecommand{\BIBforeignlanguage}[2]{{%
\expandafter\ifx\csname l@#1\endcsname\relax
\typeout{** WARNING: IEEEtran.bst: No hyphenation pattern has been}%
\typeout{** loaded for the language `#1'. Using the pattern for}%
\typeout{** the default language instead.}%
\else
\language=\csname l@#1\endcsname
\fi
#2}}
\providecommand{\BIBdecl}{\relax}
\BIBdecl

\bibitem{faultTolerance}
\BIBentryALTinterwordspacing
R.~Guerraoui and A.~Schiper, ``Fault-tolerance by replication in distributed
  systems,'' in \emph{Reliable Software Technologies — Ada-Europe '96}, ser.
  Lecture Notes in Computer Science, A.~Strohmeier, Ed.\hskip 1em plus 0.5em
  minus 0.4em\relax Springer Berlin Heidelberg, 1996, vol. 1088, pp. 38--57.
  [Online]. Available: \url{http://dx.doi.org/10.1007/BFb0013477}
\BIBentrySTDinterwordspacing

\bibitem{bookIntroduction}
C.~Cachin, R.~Guerraoui, and L.~Rodrigues, \emph{Introduction to reliable and
  secure distributed programming}.\hskip 1em plus 0.5em minus 0.4em\relax
  Springer, 2011.

\bibitem{schneider}
\BIBentryALTinterwordspacing
F.~B. Schneider, ``Implementing fault-tolerant services using the state machine
  approach: A tutorial,'' \emph{ACM Comput. Surv.}, vol.~22, no.~4, pp.
  299--319, Dec. 1990. [Online]. Available:
  \url{http://doi.acm.org/10.1145/98163.98167}
\BIBentrySTDinterwordspacing

\bibitem{lamport}
\BIBentryALTinterwordspacing
L.~Lamport, ``The part-time parliament,'' \emph{ACM Trans. Comput. Syst.},
  vol.~16, no.~2, pp. 133--169, May 1998. [Online]. Available:
  \url{http://doi.acm.org/10.1145/279227.279229}
\BIBentrySTDinterwordspacing

\bibitem{chubby}
\BIBentryALTinterwordspacing
M.~Burrows, ``The chubby lock service for loosely-coupled distributed
  systems,'' in \emph{Proceedings of the 7th Symposium on Operating Systems
  Design and Implementation}, ser. OSDI '06.\hskip 1em plus 0.5em minus
  0.4em\relax Berkeley, CA, USA: USENIX Association, 2006, pp. 335--350.
  [Online]. Available: \url{http://dl.acm.org/citation.cfm?id=1298455.1298487}
\BIBentrySTDinterwordspacing

\bibitem{viewstamp}
\BIBentryALTinterwordspacing
B.~M. Oki and B.~H. Liskov, ``Viewstamped replication: A new primary copy
  method to support highly-available distributed systems,'' in
  \emph{Proceedings of the Seventh Annual ACM Symposium on Principles of
  Distributed Computing}, ser. PODC '88.\hskip 1em plus 0.5em minus 0.4em\relax
  New York, NY, USA: ACM, 1988, pp. 8--17. [Online]. Available:
  \url{http://doi.acm.org/10.1145/62546.62549}
\BIBentrySTDinterwordspacing

\bibitem{viveladifference}
R.~van Renesse, N.~Schiper, and F.~Schneider, ``Vive la diff{\'e}rence: Paxos
  vs. viewstamped replication vs. zab,'' \emph{Dependable and Secure Computing,
  IEEE Transactions on}, vol.~PP, no.~99, pp. 1--1, 2014.

\bibitem{refinement}
\BIBentryALTinterwordspacing
L.~Lamport, ``Byzantizing paxos by refinement,'' in \emph{Proceedings of the
  25th International Conference on Distributed Computing}, ser. DISC'11.\hskip
  1em plus 0.5em minus 0.4em\relax Berlin, Heidelberg: Springer-Verlag, 2011,
  pp. 211--224. [Online]. Available:
  \url{http://dl.acm.org/citation.cfm?id=2075029.2075058}
\BIBentrySTDinterwordspacing

\bibitem{castroPracticalByzantine}
\BIBentryALTinterwordspacing
M.~Castro and B.~Liskov, ``Practical byzantine fault tolerance and proactive
  recovery,'' \emph{ACM Trans. Comput. Syst.}, vol.~20, no.~4, pp. 398--461,
  Nov. 2002. [Online]. Available:
  \url{http://doi.acm.org/10.1145/571637.571640}
\BIBentrySTDinterwordspacing

\bibitem{datacenterNonMalicious}
\BIBentryALTinterwordspacing
P.~Bhatotia, A.~Wieder, R.~Rodrigues, F.~Junqueira, and B.~Reed, ``Reliable
  data-center scale computations,'' in \emph{Proceedings of the 4th
  International Workshop on Large Scale Distributed Systems and Middleware},
  ser. LADIS '10.\hskip 1em plus 0.5em minus 0.4em\relax New York, NY, USA:
  ACM, 2010, pp. 1--6. [Online]. Available:
  \url{http://doi.acm.org/10.1145/1859184.1859186}
\BIBentrySTDinterwordspacing

\bibitem{nonMalicious}
D.~Behrens, S.~Weigert, and C.~Fetzer, ``Automatically tolerating arbitrary
  faults in non-malicious settings,'' in \emph{Dependable Computing (LADC),
  2013 Sixth Latin-American Symposium on}, April 2013, pp. 114--123.

\bibitem{hardening}
M.~Correia, D.~G. Ferro, F.~P. Junqueira, and M.~Serafini, ``Practical
  hardening of crash-tolerant systems.'' in \emph{USENIX Annual Technical
  Conference}, 2012, pp. 453--466.

\bibitem{paxosMadeLive}
\BIBentryALTinterwordspacing
T.~D. Chandra, R.~Griesemer, and J.~Redstone, ``Paxos made live: An engineering
  perspective,'' in \emph{Proceedings of the Twenty-sixth Annual ACM Symposium
  on Principles of Distributed Computing}, ser. PODC '07.\hskip 1em plus 0.5em
  minus 0.4em\relax New York, NY, USA: ACM, 2007, pp. 398--407. [Online].
  Available: \url{http://doi.acm.org/10.1145/1281100.1281103}
\BIBentrySTDinterwordspacing

\bibitem{vieira08a}
G.~M.~D. Vieira and L.~E. Buzato, ``Treplica: ubiquitous replication,'' in
  \emph{SBRC’08: Proc. of the 26th Brazilian Symposium on Computer Networks
  and Distributed Systems}, 2008.

\bibitem{vieira-tr10b}
------, ``Implementation of an object-oriented specification for active
  replication using consensus.''\hskip 1em plus 0.5em minus 0.4em\relax
  Technical Report IC-10-26, Institute of Computing, University of Campinas,
  2010.

\bibitem{generals}
\BIBentryALTinterwordspacing
L.~Lamport, R.~Shostak, and M.~Pease, ``The byzantine generals problem,''
  \emph{ACM Trans. Program. Lang. Syst.}, vol.~4, no.~3, pp. 382--401, Jul.
  1982. [Online]. Available: \url{http://doi.acm.org/10.1145/357172.357176}
\BIBentrySTDinterwordspacing

\bibitem{scrubbing}
T.~Schwarz, Q.~Xin, E.~Miller, D.~D.~E. Long, A.~Hospodor, and S.~Ng, ``Disk
  scrubbing in large archival storage systems,'' in \emph{Modeling, Analysis,
  and Simulation of Computer and Telecommunications Systems, 2004. (MASCOTS
  2004). Proceedings. The IEEE Computer Society's 12th Annual International
  Symposium on}, Oct 2004, pp. 409--418.

\bibitem{memoryIntegrity}
\BIBentryALTinterwordspacing
D.~Clarke, S.~Devadas, M.~Dijk, B.~Gassend, and G.~Suh, ``Incremental multiset
  hash functions and their application to memory integrity checking,'' in
  \emph{Advances in Cryptology - ASIACRYPT 2003}, ser. Lecture Notes in
  Computer Science, C.-S. Laih, Ed.\hskip 1em plus 0.5em minus 0.4em\relax
  Springer Berlin Heidelberg, 2003, vol. 2894, pp. 188--207. [Online].
  Available: \url{http://dx.doi.org/10.1007/978-3-540-40061-5\_12}
\BIBentrySTDinterwordspacing

\end{thebibliography}
%\begin{thebibliography}{1}


%\end{thebibliography}

%\begin{biography}[{\includegraphics[width=1in,height=1.25in,clip,keepaspectratio]{mshell}}]{Michael Shell}

%\begin{IEEEbiography}[{\includegraphics[width=1in,height=1.25in,clip,keepaspectratio]{picture}}]%{John Doe}
%\blindtext
%\end{IEEEbiography}


%\vfill


\end{document}